# Static critical exponents of the ferromagnetic transition in spin glass re-entrant systems

**Cláudia M. Haetinger[1], Luis Ghivelder[2], Jacob Schaf[1], Paulo Pureur[1]**

[1]Instituto de Física, Universidade Federal do Rio Grande do Sul, P.O. Box 15051, 91501-970 Porto Alegre, RS, Brazil

[2]Instituto de Física, Universidade Federal do Rio de Janeiro, Rio de Janeiro, P.O. Box 68528, 21945-970 Rio de Janeiro, RJ, Brazil

E-mail : ppureur@if.ufrgs.br

**Abstract**

The static critical phenomenology near the Curie temperature of the re-entrant metallic alloys $Au_{0.81}Fe_{0.19}$, $Ni_{0.78}Mn_{0.22}$, $Ni_{0.79}Mn_{0.21}$ and amorphous $a$-$Fe_{0.98}Zr_{0.08}$ is studied using a variety of experimental techniques and methods of analysis. We have generally found that the values for the exponents $\alpha$, $\beta$, $\gamma$ and $\delta$ depart significantly from the predictions for the 3D Heisenberg model and are intermediate between these expectations and the values characterizing a typical spin glass transition. Comparing the exponents obtained in our work with indices for other re-entrant systems reported in the literature, a weak universality class may be defined where the exponents distribute within a certain range around average values.

## 1. Introduction

The magnetic properties of random solid solutions have been extensively studied in the last thirty years. In a certain number of these systems, including metals and insulators, a re-entrant behavior that shares properties of both spin glass and ferromagnetic orderings have been observed [1]. When cooled from the high temperature paramagnetic phase, the re-entrant systems first enters into a ferromagnetic-like state where the spontaneous magnetization rises rapidly to large values below the Curie temperature, $T_c$ [2, 3]. However, when the temperature is decreased below a certain characteristic value $T_K$, the magnetic response of these systems displays features of spin glasses. In particular, the initial DC magnetic susceptibility shows a marked fall from the limiting value imposed by the demagnetizing effect at $T_K$ [4, 5]. Moreover, significant ZFC-FC magnetization irreversibilities are usually observed in the re-entrant region when measurements are performed with low and moderate fields [2, 3]. Anomalies reminding the spin glass state are also observed below $T_K$ in neutron diffraction spectra [6], electron paramagnetic resonance [7], Mössbauer effect [8], magneto-resistance [9], among other properties.

At first sight, the occurrence of the re-entrant phase in certain disordered magnets seems paradoxical since it suggests that the fundamental state has larger entropy than the intermediate ferromagnetic-like state. Thus, it is not surprising that much controversy and seemly contradictory results about the nature of the re-entrant as well as the ferromagnetic-like states are encountered in the literature [1]. For instance, in certain systems, such as the semiconductor $Eu_xSr_{1-x}S$, neutron diffraction studies do not show evidence for magnetic Braag scattering in the re-entrant state [10]. However, direct observation of domain structure by Lorentz microscopy in polycrystalline $Ni_{1-x}Mn_x$, and in the amorphous $(Fe_{78}Mn_{22})_{75}P_{16}B_6Al_3$ and $a$-$Fe_{1-x}Zr_x$ does not reveal significant change upon variation of the temperature through $T_K$ [11]. On the other hand, several reports focus on the qualitative differences between the ferromagnetic-like phase in the re-entrants and the conventional collinear ferromagnetic state. Examples are spin-wave excitations [12], magnetic relaxation [13] and critical properties near $T_c$ [14-17]. The re-entrant magnetic behavior is now being related to the inverse freezing problem [18] which is a current theoretical challenge aimed to describe unconventional transitions where the low temperature phase looks more entropic than the state at higher temperatures [19].

In this article, we report on an experimental study of the static critical phenomenology near $T_c$ of a number of metallic re-entrant magnets. The investigated systems are the *fcc* alloys $Au_{0.81}Fe_{0.19}$, $Ni_{0.79}Mn_{0.21}$ and $Ni_{0.78}Mn_{0.22}$. Results are also presented on the metallic amorphous $Fe_{0.92}Zr_{0.08}$. Magnetization at low fields, AC susceptibility, resistivity and specific heat measurements were performed in order to determine the critical exponents $\alpha$, $\beta$, $\gamma$ and $\delta$. Different methods were employed to analyze the results. We have generally found exponents whose values are intermediate between those largely accepted for Heisenberg ferromagnets in the ordered case [20] and those usually observed in purely spin glass systems [21]. The obtained exponents are weakly dependent on the studied system. This fact suggests that a specific universality class may not exist for the re-entrant magnetic systems. We are thus lead to suppose that although disorder is indeed relevant to the ferromagnetic transition of the re-entrant magnets, its influence on criticality may differ slightly from system to system.

Our results go along the line of previous investigations on the critical behavior near $T_c$ of re-entrant systems that report on anomalous exponent values [14-17], but contrasts with results of extensive studies on amorphous collinear ferromagnets, where disorder is reported to be irrelevant for critical phenomenology [22].

## 2. Experimental

The $Au_{0.81}Fe_{0.19}$ alloy was prepared by arc melting the constituents under argon atmosphere. The purities of the starting metals were 99.998% for Au and 99.99% for Fe. The mass loss was negligible so that the nominal stoichiometry was preserved in the resulting ingot of 1.5 g weight. Part of the ingot was rolled to a slab having width of 0.22 mm, from which samples for magnetic and transport measurements were obtained. The sample for magnetization and AC susceptibility experiments has the form of a disk with diameter 4.3 mm. The sample for resistivity measurements has the form of a parallelepiped with surface $0.95 \times 0.48$ mm$^2$. Another part of the original ingot was shaped to an ellipsoidal form with weigh 867.8 mg and used for specific heat measurements. All samples were encapsulated in an evacuated quartz ampoule, then annealed in $950^0$ C during 24h. Finally a quench into water was performed as final step to prevent Fe clustering.

Two $Ni_{1-x}$-$Mn_x$ samples with concentrations x= 0.21 and x=0.22 were prepared from high purity Ni (99.999%) and freshly cleaned Mn (99.9%). These starting materials were arc-melted under Ar 0.2 bar. Samples for magnetic, resistivity and specific heat experiments were extracted from the resulting ingots similarly to the Au-Fe case. These specific samples were sealed into a quartz tube, then annealed in vacuum at $900^0$C for 1 hour and subsequently quenched into water mixed with ice.

The $a$-$Fe_{0.92}Zr_{0.08}$ sample used in magnetic measurements was a small piece cut from ribbons prepared by melt-spinning as described in reference [23].

Magnetization measurements were performed using a MPMS Quantum Design SQUID magnetometer operating in the RSO mode. The magnetization *M* was recorded as a function of the temperature at fixed fields according to the zero-field cooling (ZFC) and field cooling (FC) prescriptions. Measurements of *M* versus *H* at fixed temperature were also done. The magnetic field magnitude in our experiments was restricted to the range 0-500 Oe. In the magnetic measurements, the field was always kept parallel to the plane of the disk-type samples in order to minimize the demagnetization effects.

The temperature dependent real and imaginary parts of the AC susceptibility were measured in the frequency range 100-6000 Hz with a Quantum Design PPMS platform. The amplitude of the exciting AC field was kept fixed to 10 Oe. No external DC field was applied in these experiments.

Accurate resistivity measurements were performed using a low frequency AC technique that employs a variable decade transformer in a compensating circuit and a lock-in amplifier as a null detector. Measurements were done in a large temperature interval encompassing the Curie temperature of the Au-Fe and Ni-Mn alloys. Temperatures were determined with a Pt sensor having 1 mK accuracy. A large number of resistivity versus *T* data points were recorded while slowly varying the temperature, so that the temperature derivative of the resistivity, $d\rho/dT$, could be numerically calculated.

The specific heat results were obtained with a quasi-adiabatic pulse technique. The addenda heat capacity was measured separately and subtracted from the data. The temperature sensor was a grounded carbon resistor recalibrated at each run. The temperatures could be determined with accuracy better than 0.1 K. The investigated temperature range extends from around 10 K to near room temperature for the Au-Fe and Ni-Mn systems. The temperature increments used in the specific heat measurements varied from 0.2 K near $T_c$ to 2 K far from this point.

## 3. Results

### 3.1 Au-Fe

Figure 1(a) shows a representative *M* versus *T* measurement for the $Au_{0.81}Fe_{0.19}$ alloy measured with *H* = 30 Oe. A clear ZFC-FC splitting occurs at low temperatures as expected for a re-entrant system. This splitting is usually associated to the canting temperature $T_K$. In figure 1(b) the magnetization is plotted as a function of the external field in several temperatures around $T_c$. From the straight line fitted to the *M* versus *H* data at low fields, and using the procedure described in reference [24], we deduce the demagnetization factor $\eta_l$ = 0.006, that is considered throughout the analysis of the magnetic measurements in the $Au_{0.81}Fe_{0.19}$ sample. This value for $\eta_l$ is in agreement with the one estimated by approximately describing the sample as an oblate ellipsoid [25].

In order to obtain a first estimate of $T_c$ and the critical exponents $\beta$ and $\gamma$, we analyze the isotherms of figure 1(b) according to the Arrot-Noakes equation of state [26],

$$(H/M)^{1/\gamma} = at + bM^{1/\beta} \quad , \tag{1}$$

where $t = (T - T_c)/T_c$ and *a*, *b* are material dependent parameters. Thus, we outline $M^{1/\beta}$ versus $(H/M)^{1/\gamma}$ plots in fixed temperatures around $T_c$ in such a way that the exponents $\beta$ and $\gamma$ could be varied until straight lines were obtained. Figure 2 shows the linearized Arrot-Noakes plots. As commonly found in the experimental studies of the critical phenomenology of disordered ferromagnetic systems, data for $Au_{0.81}Fe_{0.19}$ obtained in low applied fields deviate strongly from the straight line behavior, probably because of large domain and demagnetization effects [27]. Thus, the points represented in figure 2 were obtained in the field range 100-500 Oe. The isotherm passing through the origin defines the value of the critical temperature, $T_c$ = 177.5 K in this analysis. The corresponding critical exponents are $\gamma$ = 1.64 and $\beta$ = 0.54. Uncertainties around 10% should be considered for these parameters.

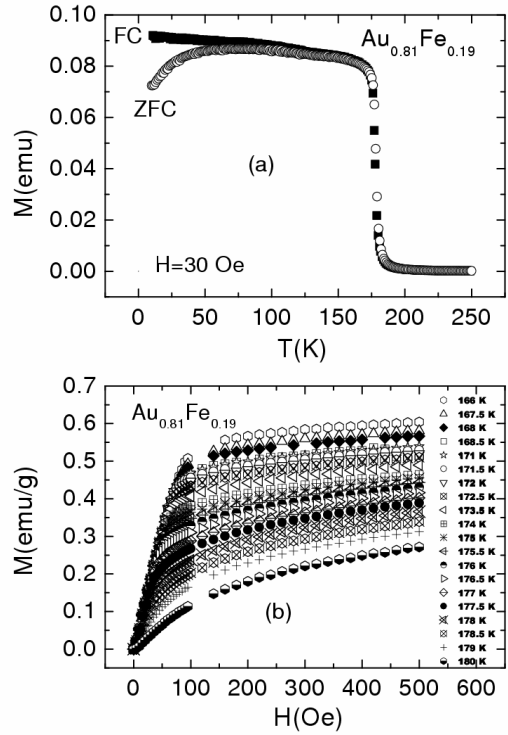

**Figure 1**. (a) Magnetic moment as a function of temperature for $Au_{0.81}Fe_{0.19}$ measured at H = 30 Oe according to the zero-field-cooling (ZFC) and field-cooling (FC) prescriptions. The Curie temperature $T_c$ is signaled. (b) Magnetization versus field in several fixed temperatures closely above and below $T_c$ for the same alloy.

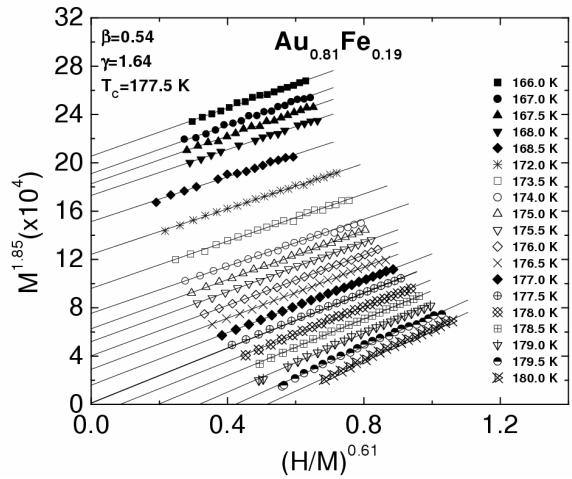

**Figure 2**. Arrot-Noakes plot for the data in Fig. 1(b). The corresponding critical exponents and $T_c$ are quoted on the figure.

The critical indices obtained with the Arrot-Noakes method were further tested based on the scaling equation of state [28],

$$M(H,t)/t^{\beta} = M(H/|t|^{\beta\delta}; \pm 1) \quad , \tag{2}$$

where the exponent $\delta$ is related to the critical isotherm and ±1 refers to temperatures above and below $T_c$, respectively. Using the scaling relation [28],

$$\beta\delta = \beta + \gamma, \qquad (3)$$

we may define the scaled magnetization $m = M(H,t)|t|^{-\beta}$ and the scaled field $h = H|t|^{-(\beta+\gamma)}$. Then, plots of $m$ versus $h$ should collapse into two universal functions $m = F_\pm(h)$, for temperatures above (+) and below (-) $T_c$. We indeed obtained good scaling of the data according to the reduced equation of state using the exponents $\beta = 0.54$ and $\gamma = 1.64$, and considering $T_c = 177$ K.

Since the critical isotherm ($t = 0$) obeys the relation [28] $M = M_0 H^{1/\delta}$, where $M_0$ is a constant, a simple plot of $\ln(M)$ as a function of $\ln H$ as the one shown in figure 3 allows the determination of $\delta$. As reported in the figure, we obtain $\delta = 4.73$ (± 0.05). This value for $\delta$ is consistent with the one calculated from equation (3) using the previously obtained values for $\gamma$ and $\beta$.

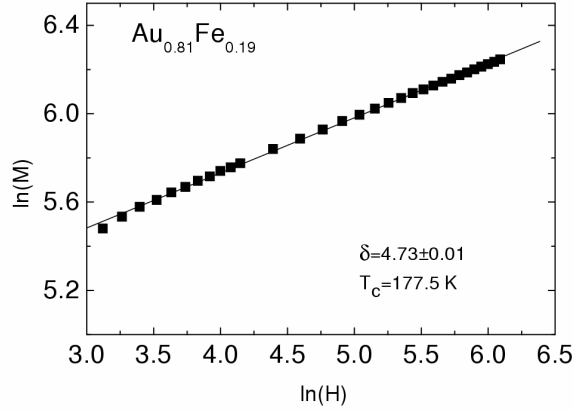

**Figure 3**. Logarithmic plot of the critical isotherm for $Au_{0.81}Fe_{0.19}$.

The critical exponents $\gamma$ and $\beta$ could be independently determined by the Kouvel-Fisher method [29]. Since the asymptotic behavior of the initial susceptibility and the spontaneous magnetization near $T_c$ are given respectively by $\chi_0 = A|t|^{-\gamma}$ and $M_s = \lim_{H \to 0} M = B|t|^\beta$, where $A$ and $B$ are constant amplitudes, one may write,

$$X(T) = -\chi_0 / (d\chi_0 / dT) = (T - T_c) / \gamma \qquad (4.a)$$

$$Y(T) = -M_s / (dM_s / dT) = (T_c - T) / \beta . \qquad (4.b)$$

Thus, the identification of linear behavior in plots of $X(T)$ vs. $T$ and $Y(T)$ vs. $T$ allows the simultaneous determination of $T_c$ and the respective critical exponents.

The use of the Kouvel-Fisher method implies the previous determination of the initial susceptibility and the spontaneous magnetization from the experimental data obtained in nonzero fields. However, instead of extrapolating the DC susceptibility data to zero field, we determined the function $X(T)$ in equation (5) directly from the measured $\chi$ in several fields between 20 and 300 Oe and judge the results on an average basis. Table 1 shows the so obtained values for $\gamma$ and $T_c$ for each applied field. We indeed found a fairly constant $\gamma$ which means that the magnetic moment is linear with the applied field as expected for a paramagnetic system in the low field range. Figure 4(a) depicts a representative Kouvel-Fisher plot for the DC susceptibility of our Au-Fe alloy. From these experiments, we obtain $\gamma = 1.63$ ($\pm\, 0.03$) and $T_c = 168 \pm 1$ K.

**Table 1**. Critical exponent $\gamma$ for the alloy $Au_{0.81}Fe_{0.19}$ obtained by applying the Kouvel-Fisher method to DC susceptibility measurements performed in the quoted fields

| System | H(Oe) | $T_c$ (K) | $\gamma$ |
|---|---|---|---|
| Au-Fe | 20 | 166.4 | 1.66 |
| | 30 | 169.1 | 1.59 |
| | 40 | 168.3 | 1.66 |
| | 50 | 166.5 | 1.65 |
| | 60 | 168.6 | 1.61 |
| | 70 | 168.6 | 1.61 |
| | 80 | 168.4 | 1.62 |
| | 90 | 168.6 | 1.61 |
| | 100 | 168.3 | 1.62 |
| | 110 | 168.6 | 1.60 |
| | 130 | 168.9 | 1.60 |
| | 150 | 166.3 | 1.65 |
| | 200 | 166.2 | 1.66 |
| | 300 | 166.1 | 1.65 |
| **Averages** | | 167.8±1.2 | 1.63±0.03 |

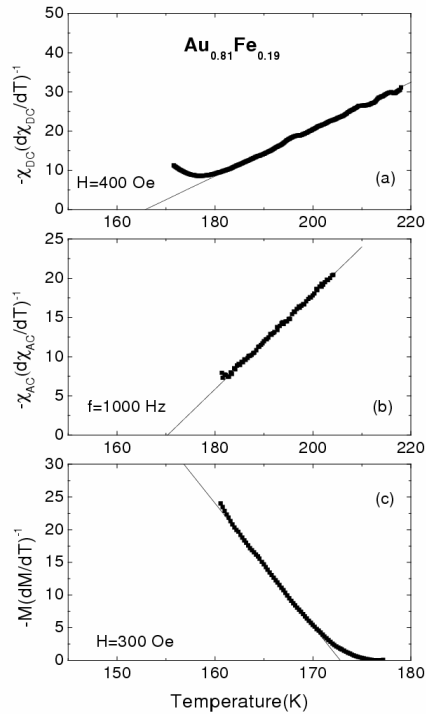

**Figure 4**. Representative Kouvel-Fisher plots of the (a) DC susceptibility, (b) AC susceptibility and (c) magnetization for $Au_{0.81}Fe_{0.19}$. The applied DC fields and employed frequency are quoted on the figures.

We found that it is crucial to take into account the demagnetizing effects for extracting meaningful results from the magnetization of the Au-Fe sample using the Kouvel-Fisher method. We thus adopt the following procedure. We select $M$ versus $T$ measurements in the range 100-300 Oe. This field range is assumed to avoid the problematic very low field data. However, the high field limit is kept as low as possible so that an excessively large distance from the critical point is prevented. Then, for each temperature we fit the $M(H)$ data to a 2$^{nd}$ order polynomial function of $H$, where the coefficients are temperature dependent. These polynomials allow us to reconstruct the $M$ versus $T$ curve for any fixed value of the internal field, $H_i = H - \eta M$, within the selected range for the applied field. In figure 4(c) we show the Kouvel-Fisher plot derived from the data corresponding to $H_i$ = 300 Oe. From this analysis we deduced $\beta$ = 0.53 and $T_c$ = 176.7 K. This value for $T_c$ is large when compared to that estimated from the paramagnetic side of the transition. AC susceptibility results are particularly useful for analysis with the Kouvel-Fischer method since they were obtained in absence of an external DC field. In order to increase the accuracy of the derived critical indices, we performed measurements in several exciting frequencies, as listed in Table 2. From those results we obtained $\gamma$ = 1.64 (± 0.02) and $T_c$ = 171.0 (± 0.6) K. Figure 4(b) shows a representative Kouvel-Fisher plot for the AC susceptibility of our Au-Fe alloy.

**Table 2**. Critical exponent $\gamma$ for the alloy $Au_{0.81}Fe_{0.19}$ obtained by applying the Kouvel-Fisher method to AC susceptibility measurements performed in the quoted frequencies. The magnitude of the AC field was 5 Oe and no DC field was superimposed.

| System | $f$ (s$^{-1}$) | $T_c$ (K) | $\gamma$ |
|---|---|---|---|
| Au-Fe | 100 | 170.8 | 1.63 |
| | 200 | 171.0 | 1.65 |
| | 300 | 172.0 | 1.62 |
| | 600 | 171.9 | 1.62 |
| | 1000 | 170.4 | 1.65 |
| | 2000 | 170.6 | 1.67 |
| | 3000 | 170.5 | 1.65 |
| | 6000 | 170.8 | 1.66 |
| Averages | | 171.0±0.6 | 1.64±0.02 |

The specific heat exponent $\alpha$ is generally the most difficult to determine in magnetic transitions of disordered systems. We perform careful specific heat measurements in the temperature interval 25-240 K, as shown by results in figure 5, but not even a feeble anomaly was observed around $T_c$.

Near to a second order magnetic phase transition the excitations contributing to the magnetic free energy and those responsible for the scattering are the same [30]. Thus, measurements of the electrical resistivity in the critical region may lead to the determination of $\alpha$. In a short temperature interval encompassing $T_c$, the temperature derivative of the resistivity may be written as [31]

$$\frac{d\rho}{dT} = \frac{C^{\pm}}{\alpha}(t^{-\alpha} - 1) + D^{\pm} \,, \qquad (5)$$

where $C$ is a critical amplitude, $D$ is a constant that measures the strength of the non-critical contribution to the resistivity and the signal ± refers to temperatures above and below $T_c$, respectively.

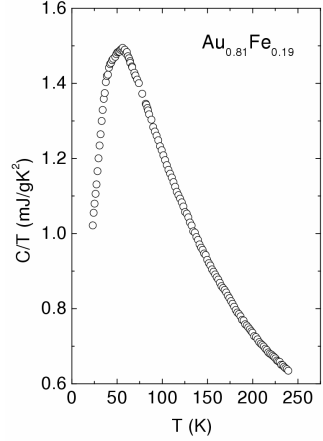

**Figure 5**. Specific heat divided by the temperature as a function of $T$ for $Au_{0.81}Fe_{0.19}$.

In spite of making repeated measurements of the Au-Fe sample resistivity in a large temperature interval around $T_c$, we could not precisely fit our results to equation (5) and were not able to extract a reliable estimate for the exponent $\alpha$ in this case. We thus deduced this exponent by assuming that the Rushbrooke,

$$\alpha + 2\beta + \gamma = 2 , \quad (6)$$

and Griffits,

$$\alpha + \beta(\delta + 1) = 2 , \quad (7)$$

scaling relations [28] are valid for the ferromagnetic transition of the re-entrant magnets. We also deduced the value for $\alpha$ in a closely related Au-Fe alloy by using the value for the correlation length exponent $v$ obtained from small angle neutron scattering experiments [32] and assuming the validity of the hyperscaling relation, $\alpha = 2 - vd$, where $d = 3$ is the dimensionality.

Table 3 condenses the values for the critical indices determined in this work and values reported in the literature for the re-entrant Au-Fe alloys. Also shown in Table 3 are the results for Ni-Mn alloys. For $Au_{0.81}Fe_{0.19}$ we obtained a good agreement in the determination of the critical exponents within the different methods used. However, the extrapolated critical temperature depends significantly on the method of analysis. In Table 3, the reported critical temperature $T_c$ is estimated from averaging the values obtained from the various methods of analysis. Previously published data on a different sample of the same alloy [17] and results reported by Gangopadhyay [15] for $Au_{0.82}Fe_{0.18}$ are also listed in Table 3. The exponents reported in reference [15] differ strongly from those obtained in the present study. Authors in [15] estimated the saturation magnetization and the initial susceptibility from their DC magnetization data from $M^3$ versus $H/M$ isotherms, then applied the Kouvel-Fisher analysis. The exponents reported by them refer to a region in the immediate vicinity of $T_c$. Their exponent $\gamma$ is found to be strongly temperature dependent and evolves to higher values when $(T - T_c)$ increases. This is unusual, since one expects that this exponent should evolve toward the mean-field expectancy $\gamma = 1$ when the temperature is progressively increased above $T_c$. In our experiments, rounding effects obscure the critical phenomenology in the immediate vicinity of $T_c$, either in the analysis of DC magnetization or AC susceptibility measurements. Thus, most probably, the exponents reported in the present investigation are related to a different reduced temperature range as the one studied in [15].

**Table 3**. Average critical exponents for the ferromagnetic transition in the $Au_{0.81}Fe_{0.19}$ alloy and closely related systems, and for the alloys $Ni_{0.78}Mn_{0.22}$ and $Ni_{0.79}Mn_{0.21}$.

| System | $\alpha$ [a] | | | $\beta$ | $\gamma$ | $\delta$ | $T_c$ (K) | Reference |
|---|---|---|---|---|---|---|---|---|
| | $\rho$ | R | G | | | | | |
| $Au_{0.81}Fe_{0.19}$ | | -0.7 | -1.1 | 0.54±0.05 | 1.64±0.02 | 4.73 | 174±4 | This work |
| $Au_{0.81}Fe_{0.19}$ | | -0.7 | -1.0 | 0.52±0.04 | 1.63±0.04 | 4.69 | 177 | [17] |
| $Au_{0.82}Fe_{0.18}$ | | -0.05 | -0.3 | 0.46±0.03 | 1.13±0.04 | 4.0±0.1 | 154 | [15] ([b]) |
| Au-Fe | -1±0.1 | | | | 2±0.2 | | | [32] ([c]) |
| $Ni_{0.78}Mn_{0.22}$ | -0.8 | -0.8 | -1.08 | 0.55±0.05 | 1.71±0.03 | 4.61±0.03 | 227±3 | This work |
| $Ni_{0.78}Mn_{0.22}$ | -0.81 | -0.79 | | | 0.54±0.04 | 1.72±0.04 | | [34] |
| $Ni_{0.79}Mn_{0.21}$ | | | | | 1.71±0.1 | | 281±3 | This work([d]) |

([a]) The values for $\alpha$ were obtained from $d\rho/dT$ results or deduced from the Rushbrooke (R) or Griffits (G) scaling relations.
([b]) $Au_{0.82}Fe_{0.18}$. Exponents obtained from Kouvel-Fisher analysis from DC magnetization measurements where the saturation magnetization and the initial susceptibility where previously determined.
([c]) $Au_{0.82}Fe_{0.18}$ and $Au_{0.80}Fe_{0.20}$. Exponent deduced from the scaling relations, $\alpha = 2 - \nu d$, and $\gamma = 2\nu$, where $\nu$ was extracted from SANS experiments.
([d]) Kouvel-Fisher analysis of AC susceptibility measurements.

*3.2 Ni-Mn*

We repeat the above reported experiments and analyses for the re-entrant magnets $Ni_{0.78}Mn_{0.22}$ and $Ni_{0.79}Mn_{0.21}$. Figure 6(a) shows a representative $M$ versus $T$ measurement for the $N_{0.78}Mn_{0.22}$ alloy measured in $H = 30$ Oe. A clear ZFC-FC splitting occurs below $T_K \approx 50$ K, where the system enters the spin glass phase. In panel (b) of figure 6 the magnetization is plotted as a function of the external field in several temperatures near $T_c$. As for the Au-Fe case, we deduced the demagnetization factor $\eta_2 = 0.062$ from the straight line fitted to the $M$ versus $H$ data in low fields. Exactly the same procedure was done for the sample $Ni_{0.79}Mn_{0.21}$. In this case the obtained demagnetization factor is $\eta_3 = 0.030$.

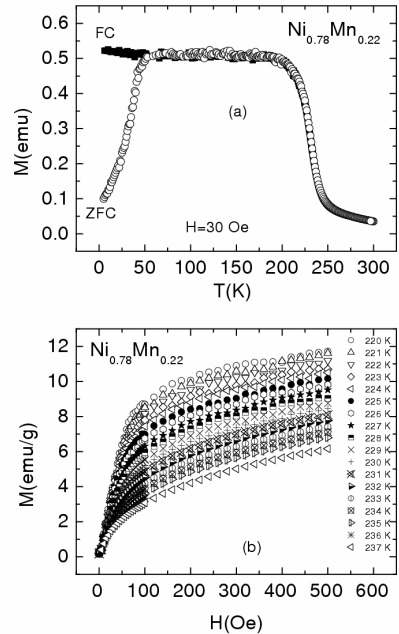

**Figure 6**. The same as Fig.1, but for $Ni_{0.78}Ni_{0.22}$

From the Arrot-Noakes analyses we obtained the exponents $\gamma = 1.71$ and $\beta = 0.55$ for $Ni_{0.78}Mn_{0.22}$ system. The same exponents were obtained in $Ni_{0.79}Mn_{0.21}$. However, this case is experimentally less clear cut and the reported exponents should be taken as less accurate estimations. Figure 7 shows the Arrot-Noakes plots for $Ni_{0.78}Mn_{0.22}$. These exponents were tested with the scaling equation of state (2) assuming the validity of relation (3). Plots of the reduced magnetization $m$ versus the scaled field $h$ collapse fairly well into two universal functions $m = F_\pm(h)$, for temperatures above (+) and below (-) $T_c$, confirming the values previously obtained for $\gamma$ and $\beta$ in both alloys.

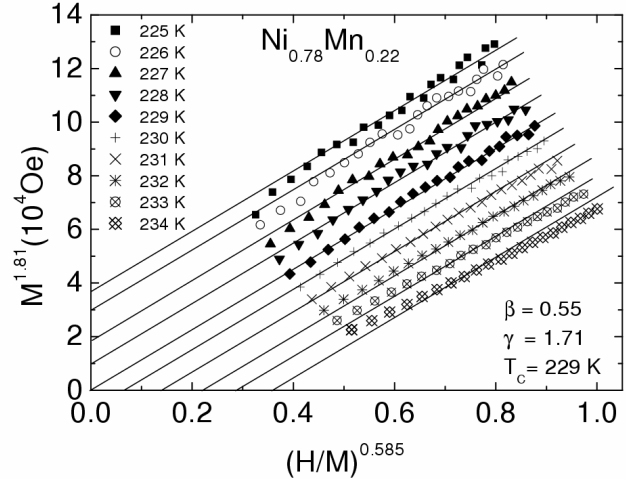

**Figure 7**. Arrot-Noakes plot for the data in Fig. 6(b). The corresponding critical exponents and $T_c$ are quoted on the figure.

Logarithmic plots of the critical magnetization isotherm as a function of the applied field allowed the extraction of the exponent $\delta = 4.61$ for $Ni_{0.78}Mn_{0.22}$.

Kouvel-Fisher analysis based on equations (4.a) and (4.b) were also performed for the Ni-Mn systems using DC and AC susceptibility measurements. The DC measurements in the paramagnetic phase were performed in several applied fields and the results for the exponent $\gamma$ were judged on an average basis. For the $Ni_{0.78}Mn_{0.22}$ alloy useful DC susceptibility results were obtained in the field range between 60 Oe and 400 Oe. The average was performed over 6 different applied fields and we estimate $\gamma = 1.71$ and $T_c = 225$ K. This value for $T_c$ is significantly smaller than that derived from the Arrot-Noakes method, but is still within the estimated range of inaccuracy. The AC susceptibility experiments performed for the $Ni_{0.78}Mn_{0.22}$ system are consistent with $\gamma = 1.71$, but the error could not be confidently estimated. For the $Ni_{0.79}Mn_{0.21}$ system, seven independent AC susceptibility measurements in the frequency range between 100 and 6000 Hz were performed. A representative Kouvel-Fisher plot of these AC susceptibility measurements is shown in figure 8. The average parameters obtained from the Kouvel-Fisher analysis in this case are $\gamma = 1.71$ and $T_c = 281$ K. The imprecision in the determination of the Curie temperature of the $Ni_{0.79}Mn_{0.21}$ is also appreciable, corroborating the difficulties for studying the critical phenomenology in the Ni-Mn system. It is interesting that in higher temperatures the AC susceptibility behaves as a power law with exponent $\gamma \approx 0.8$ in a large temperature interval above $T_c$, as may be seen in figure 8. Small susceptibility exponents have been identified above the ferromagnetic transition of semi-disordered systems with spinel structure [33]. In the case of the $Ni_{0.79}Mn_{0.21}$ alloy, the regime with $\gamma$ smaller than the mean-field value extrapolates to a too much high critical temperature, so that it can not be considered as an asymptotic behavior.

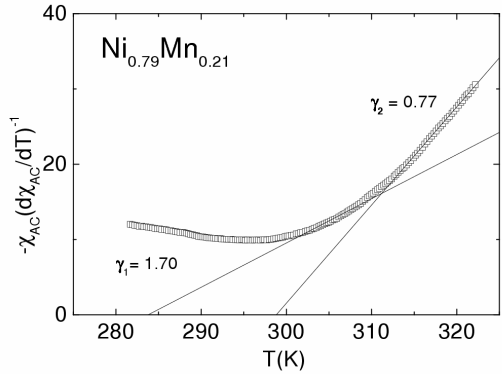

**Figure 8**. Representative Kouvel-Fischer plot for the AC susceptibility of $Ni_{0.79}Fe_{0.21}$. The applied frequency and exponents are quoted on the figure.

Specific heat measurements (not shown) were performed for the $Ni_{0.78}Mn_{0.22}$ re-entrant system. No anomaly could be seen near the Curie temperature. However, fits of $d\rho/dT$ results to equation (5) allowed a rough estimation of the critical exponent $\alpha$ for this alloy in the paramagnetic side of the transition. A representative measurement of $d\rho/dT$ for the $Ni_{0.78}Mn_{0.22}$ alloy is shown in figure 9. The critical parameters corresponding the continuous fitting line are $\alpha = -0.8 \pm 0.1$ and $T_c = 229\ (\pm 1)$ K. The values for $\alpha$ and $T_c$ are coincident with those previously obtained using resistivity measurements in an alloy of the same composition [34]. In the magnetically ordered state, a fit of $d\rho/dT$ to equation (5) could not be done because of an interesting peculiarity of the resistivity of $Ni_{0.78}Mn_{0.22}$ occurring just below the Curie temperature. As shown in figure 10, a maximum reminiscent of the opening of a super-zone gap in the Fermi surface is clearly evidenced in measurements performed at zero and low applied fields. Fields above 50 Oe applied parallel to the current strongly suppress the effect, that is completely removed at 100 Oe. The observation of a super-zone effect in the resistivity means that in low applied fields an antiferromagnetic ordering competes with and becomes favorable over the ferromagnetic coupling in temperatures nearly below $T_c$. This effect should occur in some regions of the sample having the size of the electron mean-free path or larger. The same effect was also observed in the resistivity of some samples of the Heusler compound $Pd_2MnSn$, where antiferromagnetic coupling was suggested to dominate over the ferromagnetic ordering in spatially limited regions in temperatures closely below $T_c$ [34].

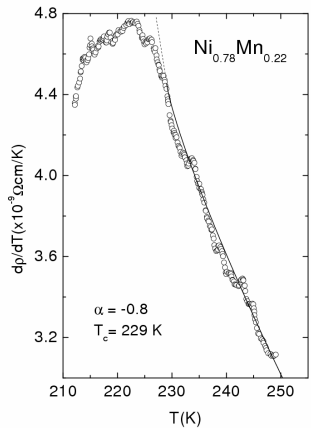

**Figure 9**. Temperature derivative of the resistivity versus T for $Ni_{0.78}Ni_{0.22}$. The continuous line corresponds to fit to Eq. (5) in the range $T > T_c$. The relevant critical exponent is quoted.

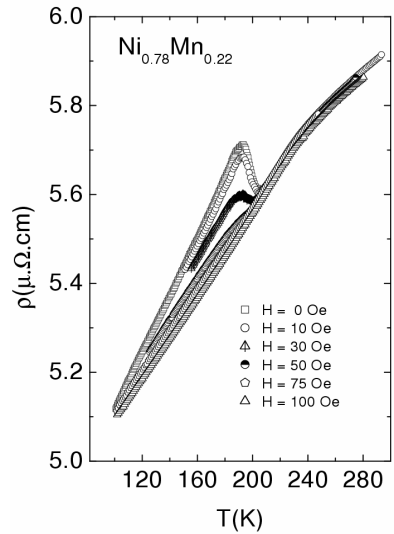

**Figure 10.** Resistivity versus temperature near $T_c$ for $Ni_{0.78}Ni_{0.22}$. The quoted fields were applied parallel to the current.

Table 3 lists the critical indices obtained for the Ni-Mn alloys. As for $Au_{0.81}Fe_{0.19}$, reported exponents and respective errors are averages over the values obtained from each experiment and method used to analyze the results.

*3.3 Fe-Zr*

We investigated the magnetization and AC susceptibility of the amorphous alloy $Fe_{0.92}Zr_{0.08}$ and applied the Kouvel-Fisher method to extract the $\gamma$ and $\beta$ exponents. A representative analysis of the AC susceptibility measurements is shown in figure 11. Using the average process of several DC susceptibility experiments in several fields below 100 Oe, we obtained $\gamma = 1.75 \pm 0.03$ and $T_c = 187 (\pm 1)$ K, whereas from the average of AC susceptibility measurements performed in different frequencies, we derived $\gamma = 1.76 \pm 0.02$ and $T_c = 184 (\pm 3)$ K. The analysis of the magnetization near the Curie temperature was difficult in spite of the negligible demagnetization factor of the thin amorphous tape oriented along the field. The value obtained is $\beta = 0.66 (\pm 0.06)$.

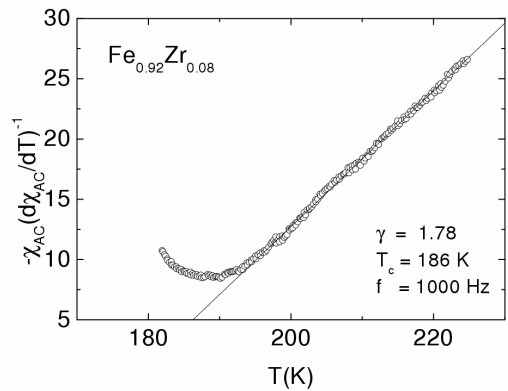

**Figure 11.** Representative Kouvel-Fischer plot for the AC susceptibility of $a$-$Fe_{0.92}Zr_{0.08}$. The applied frequency and the susceptibility exponent are quoted on the figure.

A controversy exists about the static critical exponents in *a*-FeZr alloys. Authors in refs. [14, 34-37] found values for $\alpha$, $\beta$, $\gamma$, and $\delta$ substantially higher than expected for ordered ferromagnets. On the other hand, Kaul [22, 38], Reisser et al. [27] and Ma et al. [39] report on the ferromagnetic transition of a number of *a*-FeZr with near Heisenberg-like $\beta$, $\gamma$, and $\delta$ exponents. Table 4 shows the anomalous exponents found for the *a*-FeZr systems. A significant dispersion occurs among these indices, even for alloys having the same composition. This fact is a further indication that the critical behavior in these amorphous ferromagnets is strongly sample-dependent. Some alloys seem representative of re-entrant magnets, where disorder is non-trivial (accompanied by frustration), while others reproduce the critical phenomenology of the ordered case, indicating that disorder in these cases is irrelevant.

**Table 4**. Anomalous critical exponents for the ferromagnetic transition in *a*-FeZr.

| Alloy | $\alpha$ | $\beta$ | $\gamma$ | $\delta$ | $T_c$ (K) | Reference |
|---|---|---|---|---|---|---|
| $Fe_{0.92}Zr_{0.08}$ |  | 0.62 | 1.92 | 5.82 | 174.6 | [14] |
|  | -1.1 |  |  |  |  | [36] |
|  |  | 0.66 | 1.76 |  | 186 | This work |
| $Fe_{0.90}Zr_{0.10}$ |  | 0.56 | 1.87 | 4.84 | 227.6 | [14] |
|  | -0.68 | 0.44 | 1.79 | 5.10 | 230 | [37] |
| $Fe_{0.895}Zr_{0.105}$ | -0.93 | 0.47 | 2.00 | 5.31 | 224 | [37] |
| **Averages** | **-0.90±0.2** | **0.55±0.1** | **1.87±0.1** | **5.30±0.5** |  |  |

## 4. Discussion

Table 5 condenses the main results on the static critical exponents for the ferromagnetic transition of the re-entrant alloys studied by us and values encountered in the literature for the same or closely related systems. Also listed are exponents experimentally found for other re-entrant systems. Because of the controversy on the critical phenomenology for the *a*-FeZr alloys, we did not include these results in Table 5. In order to allow comparisons, however, we list the exponents for some classical crystalline ferromagnets [22] and for a typical spin glass [40], as well as the most accepted theoretical expectations for these indices in the ordered case [20] and disordered cases [41]. A theoretical study of the critical phenomenology in the specific case of the re-entrant magnets is still lacking. Listed in Table 5 are the asymptotic values reported by Sobotta and Wagner [41], who proposed a renormalization-group calculation of the static critical behavior in highly disordered ferromagnets in the limit of small concentration of magnetic atoms.

With few exceptions, one observes that the values for the static exponents for the re-entrant systems are very different from those observed and predicted in the ordered ferromagnets. From data in Table 5 one thus concludes that disorder is indeed relevant in the critical phenomenology related to the ferromagnetic transition in the re-entrant magnets. This fact is in contrast with the behavior of collinear amorphous ferromagnets, where the measured exponents have values close to that of ordered Heisenberg systems [22].

Inspection of Table 5 reveals that the values for $\alpha$, $\beta$ and $\gamma$ in the re-entrant systems are systematically in-between the values that describe the phase transition of classical three-dimensional ferromagnetic materials and those characterizing a spin glass transition.

**Table 5**. Anomalous critical exponents obtained experimentally for the ferromagnetic transition in several re-entrant systems. Parameters for other relevant systems are also listed for comparison.

| System | $\alpha$ | $\beta$ | $\gamma$ | $\delta$ | Reference |
|---|---|---|---|---|---|
| $Au_{0.81}Fe_{0.19}$ |  | 0.54 | 1.64 | 4.73 | This work |
| $Au_{0.81}Fe_{0.19}$ |  | 0.52 | 1.63 | 4.69 | [17] |
| $Au_{0.82}Fe_{0.18}$ | $-1^{(a)}$ |  | $2^{(a)}$ |  | [32] |
| $Ni_{0.78}Mn_{0.22}$ | $-0.8^{(b)}$ | 0.55 | 1.71 | 4.61 | This work |
| $Ni_{0.78}Mn_{0.22}$ | $-0.81^{(b)}$ | 0.54 | 1.72 |  | [32] |
| $Ni_{0.79}Mn_{0.21}$ |  |  | 1.71 |  | This work |
| $Cd(Cr_{1-x}In_x)_2S_4$ | $-1^{(a)}$ |  | 2 |  | [43] |
| $Eu_{0.7}Sr_{0.3}S$ | $-0.48^{(c)}$ |  |  |  | [44] |
| $Eu_{0.8}Sr_{0.2}S_{0.5}Se_{0.5}$ |  | 0.44 | 1.84 | 5.0 | [45] |
| (PdFe)Mn |  | 0.53 | 1.64 | 4.1 | [46] |
| **Averages** | **-0.81±0.3** | **0.52±0.05** | **1.75±0.2** | **4.6±0.4** |  |
| Ag-Mn spin glass | -2.2 | 1.0 | 2.2 | 1.4 | [40] |
| $Fe^{(d)}$ | -0.10 | 0.36 | 1.4 | 4.35 | [22] |
| $Ni^{(d)}$ | -0.09 | 0.37 | 1.32 | 4.5 | [22] |
| Theory (D) | -1 | 0.5 | 2 | 5 | [42] |
| Theory (O) | -0.12 | 0.36 | 1.39 | 4.8 | [20] |

(a) estimated from SANS experiments and scaling relations (see text and Table 3)
(b) derived from $d\rho/dT$ measurements
(c) derived from specific heat measurements
(d) averages of values listed in reference [20]
(D) disordered case
(O) 3D Heisenberg model in the ordered case

Non-trivial disorder, associated to canting and frustration, is the distinctive feature both in the re-entrant magnets and in spin glasses. Likely, this is the origin for the non-conventional critical phenomenology in these magnetic systems. However, from the data in Table 5 we can not infer the existence of a conventional universality class for the ferromagnetic transition of the re-entrants. Given the reported experimental uncertainties for the listed exponents, one can at most estimate that a "weak universality" may exists, allowing small departures from the average values given by:

$$\alpha_m = -0.8 \ (\pm 0.3) \ ; \ \beta_m = 0.52 \ (\pm 0.05) \ ; \ \gamma_m = 1.75 \ (\pm 0.2) \ ; \ \delta_m = 4.6 \ (\pm 0.4).$$

It is noticeable that the average exponents $\alpha_m$, $\beta_m$ and $\gamma_m$ are compatible with the Rushbrooke, $\alpha_m + 2\beta_m + 2\gamma_m = 2$, and Griffiths, $\alpha_m + \beta_m(\delta_m + 1) = 2$, scaling relations. For the reported average values, the relations hold as inequalities. However, given the associated errors, the equality is also possible. The Widom, $\gamma_m = \beta_m(\delta_m - 1)$, and hyperscaling, $\alpha_m = 2 - \nu_m d$, scaling relations also are compatible with the above reported average values within the admitted uncertainty intervals. For the specific reported values, the Widom relation is slightly violated, whereas hyperscaling hold as an inequality.

The non observance of a strict universality class in the case of the ferromagnetic transition of re-entrant systems is a probable consequence of particularities in the mechanisms leading to magnetic disorder in different systems. For instance, in Au-Fe alloys a homogeneous disordered state may be obtained provided that the tendency for Fe clustering is impeded. Long-range magnetic interactions seem to prevail in this case. On the other hand, the resistivity results of figure 10 suggest that in the Ni-Mn systems, magnetic disorder is related to a subtle phase separation where antiferromagnetic regions having at least the size of the electron mean-free-path nucleate inside the ferromagnetic background. In such a system, the canting must occur mainly at the boundaries between the ferro and antiferromagnetic separated phases. Probably, short-range interactions are more relevant to explain the macroscopic re-entrant behavior in this case. The elusive criticality of $a$-FeZr also migth be related to some sample-dependent canting mechanism and to the balance between the role of short- and long-range spin interactions.

From the calculations by Sobotta and Wagner [41, 42] for randomly quenched ferromagnets one should expect the values $\alpha \rightarrow -1$, $\beta \rightarrow 0.5$, $\gamma \rightarrow 2$, and $\delta \rightarrow 5$. Although these estimates do not fit exactly the exponents measured in the present study and the average values of Table 5, it is clear that the theoretical predictions are consistent with the experimental tendency. Unfortunately, more accurate calculations of the critical exponents and a definitive answer to the question of the existence or not of universality classes for the ferromagnetic transition in the re-entrants systems, and in disordered ferromagnets in general, are still not available for the time being.

## 5. Conclusions

We have studied the critical phenomenology near the paramagnetic-ferromagnetic transition of the disordered re-entrant magnetic alloys $Au_{0.81}Fe_{0.19}$, $Ni_{0.88}Mn_{0.22}$, $Ni_{0.89}Mn_{0.21}$, and the amorphous $a$-$Fe_{0.92}Zr_{0.08}$. Using several experimental techniques and different methods for analyzing the results, we were able to obtain the static critical exponents $\alpha$, $\beta$, $\gamma$ and $\delta$ in most cases. The values found for these exponents are between those observed in a typical spin-glass transition and the expectation for a classical ferromagnetic transition. This finding contrasts to the widely studied and much better understood situation of amorphous ferromagnets with collinear spins. In the latter, disorder was shown to be non-relevant and the critical exponents are the same as those for the ordered systems [22].

The problem of the influence of non-trivial disorder, which is associated to canting and frustration in the critical phenomenology of the ferromagnetic transition is scarcely studied, either experimentally as theoretically. Some recent efforts to systematize the critical behavior of systems with different degrees of disorder clearly show the difficulties to drawn a general picture about this subject [33]. Our results fits into a rough systematic represented by the results in Table 5 for a number of re-entrant magnets, both metallic and insulating, that seems to define a weak universality class where exponents are distributed within significant intervals around average values. However, a true universality class describing a unique critical phenomenology near the Curie temperature of the re-entrant systems is probably inexistent because of the various microscopic mechanisms leading to spin-disorder in different systems. A relevant result of our work related this issue is the observation of super-zone effects in the resistivity near the ferromagnetic transition of $Ni_{0.88}Mn_{0.22}$. As a general conclusion, we have found that the non-trivial spin disorder characteristic of the re-entrant systems leads to critical exponents that are differs significantly different from those of ordered ferromagnets.

**Acknowledgements**. This work was partially supported by the Brazilian Agencies CNPq and FAPERGS under the grant PRONEX 04.0938.0. LG is also supported by FAPERJ.